\documentclass[12pt,a4paper]{article}
\usepackage[a4paper, total={6.5in, 9.5in}]{geometry}

\usepackage{amsthm,amsmath,amssymb,amsfonts,bm}
\usepackage{graphicx}
\usepackage{placeins}
\usepackage{epstopdf}
\usepackage{multirow}

\def\hang{\hangindent\parindent}
\def\rf{\par\noindent\hang}

\DeclareMathAlphabet{\mathpzc}{OT1}{pzc}{m}{it}

\begin{document}

\baselineskip=20pt

\begin{center}
%\noindent
{\bf \Large A comparison of strategies for selecting auxiliary variables for multiple imputation}
\end{center}

\bigskip

\begin{center}
{\bf Rheanna M. Mainzer*$^{1}$, Cattram D. Nguyen$^{1,2}$, John B. Carlin$^{1,3}$, Margarita Moreno-Betancur$^{1}$, Ian R. White$^{4}$, Katherine J. Lee$^{1,2}$}
\end{center}

\medskip

%\begin{center}
%{
{\sl \rf \small $^{1}$Clinical Epidemiology and Biostatistics Unit, Murdoch Children's Research Institute, Parkville, Victoria 3052, Australia \\
$^{2}$Department of Paediatrics, The University of Melbourne, Parkville, Victoria 3052, Australia \\
$^{3}$Centre for Epidemiology and Biostatistics, Melbourne School of Population and Global Healh, The University of Melbourne, Parkville, Victoria 3052, Australia \\ 
$^{4}$MRC Clinical Trials Unit, University College London, 90 High Holborn, London, UK}
%}
%\end{center}

\medskip

{\rf \textbf{Correspondence}: $^*$Rheanna Mainzer, Murdoch Children's Research Institute, 50 Flemington Road, Parkville, Victoria 3052, Australia. Email: rheanna.mainzer@mcri.edu.au}

\medskip

\rf \textbf{Keywords}: Missing data, imputation model, auxiliary variables, variable selection

\newpage

\begin{center}
{\bf ABSTRACT}
\end{center}
Multiple imputation (MI) is a popular method for handling missing data. Auxiliary variables can be added to the imputation model(s) to improve MI estimates. However, the choice of which auxiliary variables to include in the imputation model is not always straightforward. Including too few may lead to important information being discarded, but including too many can cause problems with convergence of the estimation procedures for imputation models. Several data-driven auxiliary variable selection strategies have been proposed. This paper uses a simulation study and a case study to provide a comprehensive comparison of the performance of eight auxiliary variable selection strategies, with the aim of providing practical advice to users of MI. A complete case analysis and an MI analysis with all auxiliary variables included in the imputation model (the full model) were also performed for comparison. Our simulation study results suggest that the full model outperforms all auxiliary variable selection strategies, providing further support for adopting an inclusive auxiliary variable strategy where possible. 
Auxiliary variable selection using the Least Absolute Selection and Shrinkage Operator (LASSO) was the best performing auxiliary variable selection strategy overall and is a promising alternative when the full model fails.
All MI analysis strategies that we were able to apply to the case study led to similar estimates.

\newpage

\section{Introduction} \label{sec:intro}

Missing data are often encountered in medical research. 
Many studies aim to estimate an expected value (e.g. a mean or proportion) or an exposure-outcome association.
Restricting the analysis to individuals with available data, i.e., analysing ``complete cases,'' can lead to bias or loss of precision in these estimates compared to if all data were observed.\cite{little2002statistical} 
Multiple imputation (MI) is a popular two-stage method for handling missing data that can produce valid estimates and standard errors of target quantities under relaxed assumptions regarding the mechanism leading to missing data.\cite{rubin2004multiple,van2018flexible}
In the first stage of MI, missing data are imputed multiple times with random draws from the predictive distribution of the missing values given the observed data and a specified imputation model. In the second stage, the statistical analysis of interest is applied to each imputed dataset and the results are combined using Rubin’s rules to obtain a single estimate with associated standard error.\cite{rubin2004multiple} 
When and how MI should be applied depends on the target of analysis, the reasons for missing data and whether any of the missing information can be recovered.\cite{moreno2018canonical} 

MI is usually carried out using either multiple imputation by chained equations (MICE) or multivariate normal imputation (MVNI). MICE, also known as ``fully conditional specification'', ``regression switching'' or ``sequential regression multiple imputation'', is a flexible MI approach in which univariate imputation models are specified for each variable with missing data.\cite{white2011multiple} 
Imputed values for each variable with missing data are then generated using these models, one variable at a time, until all missing values are replaced (one cycle).  
The algorithm carries out a number of cycles before obtaining one imputed dataset, and then
this procedure is repeated $m$ times to obtain $m$ sets of complete data that differ in their imputed values. 
The second approach, MVNI, is based on the assumption that all incomplete variables jointly follow a multivariate normal distribution. Imputations are obtained from this model using the Data Augmentation algorithm. \cite{schafer1997analysis}

Although the availability of MI procedures in statistical software such as R, Stata and SAS has made MI widely accessible and easy to implement in practice, the user still needs to make a number of decisions in order to carry out an analysis using MI, including what variables to include in the imputation model and in what form. \cite{van1999multiple} Best practice is to include all variables that appear in the main analysis, in the same form, as predictors in the imputation model.\cite{meng1994multiple} Failure to do so could introduce bias in MI estimates. A benefit of MI is that additional variables that do not appear in the main analysis, known as ``auxiliary variables'', can be added to the imputation model to improve performance by reducing bias and/or increasing precision.\cite{collins2001comparison,schafer2003multiple} 
Candidate auxiliary variables include variables that are related to the variables with missing data and possibly also related to the missingness of the variables with missing data.\cite{schafer1997analysis}
The extent to which auxiliary variables reduce bias or improve precision in MI estimates depends on the missingness mechanism, the target estimand and the relationship between the auxiliary variables and the variables with missing values.\cite{collins2001comparison, hardt2012auxiliary}

The choice of which auxiliary variables to include in the imputation model is not always straightforward. On one hand, imputations should not be created using a model that is more restrictive than needed,\cite{van2018flexible} 
% Section 12.1.3 Some don’ts
but on the other, including too many auxiliary variables in an imputation model can lead to a number of problems in practice.\cite{carpenter2008brief} 
Imputation algorithms may automatically drop variables or fail completely due to numerical problems such as perfect prediction, where categories of the response variable are perfectly separated by a linear combination of the covariates, or collinearity, where covariates are highly correlated.\cite{nguyen2021practical,white2010avoiding}
These problems often arise in large-scale longitudinal studies where there are many potential auxiliary variables to choose from. 
Even when imputations are successfully produced, after a sufficient number of suitable auxiliary variables have been included in the imputation model the benefits of adding additional auxiliary variables may be small.\cite{graham2012missing} One study found that including too many auxiliary variables can have a detrimental effect on the bias and precision of MI estimates, which led to the recommendation that the number of variables included in the imputation model should be no more than a third of the number of cases with complete data.\cite{hardt2012auxiliary}
Others suggest that no more than 15-25 variables are needed in the imputation model in practice.\cite{van1999multiple}
These arguments motivate the desire to employ an auxiliary variable selection strategy when there are a large number of potential auxiliary variables available. 

Several strategies have been proposed to specify an imputation model for an incomplete variable using a reduced set of auxiliary variables. Van Buuren et al. propose a four-step strategy for quick, data-based selection of variables to be included in the imputation model for the incomplete variable.\cite{van2018flexible, van1999multiple} In short, these steps are as follows. First, include all variables that appear in the analysis model. Then add variables that are related to the nonresponse in the incomplete variable according to a suitable criterion. Next, add variables that explain a considerable amount of variance in the incomplete variable. Lastly, remove any variables in steps two and three that have too many missing values. Variables in steps two and three can be chosen based on their correlation with the missingness indicator of the incomplete variable and their correlation with the incomplete variable, respectively, while variables in step four can be identified by a criterion such as the proportion of usable cases (i.e., the proportion of observed cases in the variable being selected within the subset of cases where the incomplete variable is unobserved \cite{van1999multiple}).
This strategy has been implemented in the \texttt{quickpred} function in the R package \texttt{mice} to specify imputation models for each incomplete variable in the MICE procedure,\cite{buuren2010mice} and has been adopted in practice to choose an imputation model when the full model is not feasible.\cite{clark2003developing,heymans2007variable} A similar strategy is given by Graham, \cite{graham2012missing} %Page 197
with guidance on correlation cut-offs for inclusion of variables (0.4 if the variable is a cause of missingness, 0.5 otherwise, p.197) but it is noted that these can be reduced to include more variables if there are a relatively small number of variables in the analysis. Howard et al. \cite{howard2015using} suggest using principal components of auxiliary variables in the imputation model, instead of the auxiliary variables themselves.
Strictly speaking, this approach is a data reduction method rather than a variable selection method, but we included it in this comparison since it has also been adopted in practice to specify the imputation model, \cite{erentaite2018does,jensen2018parents,jensen2019sisters,latzman2015predicting,metzger2018intersection,nair2018acculturation,roche2019autonomy}
with the R package \texttt{PcAux} available to aid its implementation.\cite{lang2017pcaux}

Other strategies for selecting auxiliary variables utilise stepwise selection, t-tests or penalised regression. The \texttt{ice} package in Stata has an option to construct imputation models for each incomplete variable using stepwise selection.\cite{royston2011multiple} An additional step is added between the initialization of the MICE algorithm and performing the imputation, where the \texttt{stepwise} command is used to select variables for each of the univariate imputation models. A related approach proposed by Andridge and Thompson uses forward selection with the fraction of missing information due to nonresponse (FMI, estimated as the ratio of between-imputation variance to total variance for a given estimator \cite{little2002statistical}) for the mean of the incomplete variable as the selection criterion for inclusion of auxiliary variables in the imputation model for that variable.\cite{andridge2015using} Dixon described the implementation of two sample t-tests to identify variables that are associated with missingness in another incomplete variable.\cite{dixon1983bmdp} The sample is split into two groups based on the presence or absence of values of the incomplete variable. T-tests are then applied to the other variables using this grouping, with large test statistics providing evidence that the probability of being missing is not the same for all cases, i.e., data are not missing completely at random (MCAR).\cite{rubin1976inference} Enders suggests results from such t-tests can be used to identify, and limit the number of, potential auxiliary variables. \cite{enders2010applied} %Page 132
Another approach is to combine penalised regression with multiple imputation. For example, the Least Absolute Shrinkage and Selection Operator (LASSO) can be used to simultaneously perform variable regularisation and selection.\cite{james2013introduction} This approach has been used for dimension reduction in the context of high-dimensional data.\cite{zhao2016multiple}

The aim of this study was to provide a comprehensive comparison of strategies used to select the set of auxiliary variables for the imputation model in MI. Motivated by a case study based on the Longitudinal Study of Australian Children (LSAC), we restrict attention to the scenario where there is an incomplete continuous outcome, a completely observed continuous exposure, a set of completely observed continuous auxiliary variables and no confounding. We use a simulation study to evaluate the performance of eight available strategies for selecting auxiliary variables for an imputation model: (1, 2) two versions of the four-step strategy provided by Van Buuren et al.\cite{van2018flexible, van1999multiple}; (3) using principal components of auxiliary variables; (4, 5) forward and forward stepwise selection; (6) forward selection based on the FMI; (7) hypothesis tests of the MCAR assumption and (8) the LASSO. We compare these eight auxiliary variable selection strategies against two ``benchmark'' analysis strategies that do not use any auxiliary variable selection: complete case analysis and MI using all auxiliary variables (the ``full'' imputation model). We investigate three scenarios of particular interest, including a scenario in which the full imputation model is likely to encounter problems. We also illustrate the performance of these strategies by applying them to the motivating LSAC case study. In Section 2 we describe the motivating case study, the design of the simulation study and the auxiliary variable selection strategies that were evaluated. Results are provided in Section 3. A discussion of these results and a summary of our key messages are provided in Section 4.

\section{Methods} \label{sec:meth}

\subsection{Motivating example: The Longitudinal Study of Australian Children}
\label{sec:LSAC}

The Longitudinal Study of Australian Children (LSAC) is a large-scale longitudinal study, with the aim of investigating the effect of a child’s environment on their well-being throughout life.\cite{sanson2002introducing} Two cohorts were recruited into the study in 2003: the baby ``B'' cohort, consisting of participants aged 0-1 years at wave 1, and the kindergarten ``K'' cohort, consisting of participants aged 4-5 years at wave 1. This case study uses data from the 4983 children in the K cohort of LSAC to examine the association between body mass index (BMI) z-score at 4-5 years of age and health-related quality of life (HRQoL) problems at 12-13 years of age. This is a simplified version of a published analysis. \cite{jansen2013bidirectional} 

The analysis of interest aimed to estimate the effect of BMI z-score on HRQoL using a linear regression model with adjustment for potential confounders. The outcome of interest, \textsl{HRQoL}, was measured using the Total Scale Score of the Paediatric Quality of Life Inventory (PedsQL) collected at wave 4.\cite{varni2003pedsql} The PedsQL scale is made up of either 21 or 23 items (depending on the child’s age), each of which is measured on a 5-point Likert scale with 1 corresponding to “never a problem” and 5 corresponding to “always a problem”. The Total Scale Score is calculated as the average of the observed scale items, after a reverse linear transformation of items in the following manner: 1=100, 2=75, 3=50, 4=25 and 5=0.  The exposure was BMI z-score at wave 1. BMI at each wave was calculated using height and weight measurements taken by an interviewer. These measurements were standardised by age and sex to derive the BMI z-score. Several potentially confounding covariates were measured at wave 1: child sex, child age, whether the child was Aboriginal or Torres Strait Islander, whether the child had a non-English speaking background and the socioeconomic position of the child's family.  

The parameter of interest was the coefficient of \textsl{BMIz} in the following linear regression model, adjusted for potential confounders:
\begin{equation}
E(\textsl{HRQoL})= \beta_0 + \beta_1 \textsl{BMIz} + \beta_2 \textsl{Female} + \beta_3 \textsl{Age} + \beta_4 \textsl{IndStat} + \beta_5 \textsl{NonEng} + \beta_6 \textsl{SEP},
\end{equation}
where \textsl{Female} is an indicator for whether the child is female, \textsl{Age} is the age in months for the child at wave 1, \textsl{IndStat} is an indicator for whether the child is Aboriginal or Torres Strait Islander, \textsl{NonEng} is an indicator for whether the child has a non-English speaking background and \textsl{SEP} is a standardised score representing the socioeconomic position of the child’s family at wave 1.
The \textsl{HRQoL} outcome was missing for 17.4\% of participants. A small percentage of participants were also missing values for \textsl{BMIz}, \textsl{IndStat}, \textsl{NonEng} and \textsl{SEP} (Supporting Information Table 1). 
We have shown previously that individuals with missing data on one or more variables in the analysis model had lower mean \textsl{HRQoL} and \textsl{SEP}, and were more likely to be Aboriginal or Torres Strait Islander or come from a non-English speaking background than individuals with complete data on all variables.\cite{mainzer2021comparison} Furthermore, a large number of potential auxiliary variables were available in the dataset. The assumption that non-response in \textsl{HRQoL} does not depend on \textsl{HRQoL} itself after conditioning on variables in the imputation model (under which MI is valid) is more plausible when the conditioning set includes auxiliary variables, and even more so if there are a large number of these. Thus MI was identified as a more appropriate method of estimating $\beta_1$ than a complete case analysis, which conditions only on variables in the analysis model.
We note that it may be the case that missingness in \textsl{HRQoL} depended on \textsl{HRQoL} even after conditioning on analysis and auxiliary variables, i.e. \textsl{HRQoL} is ``missing not at random'', but we do not consider this further.

A subset of auxiliary variables for inclusion in the imputation model was initially identified across the four waves of data collection based on substantive knowledge. This subset was:
the PedsQL scale items (measured at waves 1 -- 3); Global Health Measure (GHM, measured at waves 1 -- 4), a rating of the child’s current health;  Special health care needs (SHCN, measured at waves 1 -- 4), a variable that indicates whether the child has a health condition that requires special care; Strengths and Difficulties questionnaire (SDQ, measured at waves 1 -- 4), a scale score providing information on the child’s behavior; Matrix Reasoning test (MR, measured at waves 2 -- 4), a measure of the non-verbal intelligence of the child; and the Peabody Picture Vocabulary Test (PPVT, measured at waves 1 -- 3), a scale score providing information on the child’s vocabulary. The case study had two major complexities that were not incorporated into the design of the simulation study (next section). Firstly, the exposure, two covariates and all auxiliary variables had some missing values. Secondly, the case study involved different types of variables (including highly skewed ordinal PedsQL items). Further details of the variables in this case study, including percentage of missing data, are provided in Supporting Information Table 1. 

MI was initially attempted using MICE in Stata (16.1, StataCorp LLC, College Station, TX) with $m = 20$ imputations including all seven variables in the analysis model, as well as all auxiliary variables listed above. Continuous variables (including SDQ and GHM) were imputed using linear regression, except for HRQoL, which was imputed using predictive mean matching (PMM, a semiparametric approach where values are imputed using random draws from the observed data of a set of donors with similar predicted means \cite{morris2014tuning}) with 5 donors because this variable exhibited negative skew. Binary variables were imputed using logistic regression and PedsQL items were imputed using ordinal logistic regression. The \texttt{augment} option was used to perform augmented regression in the presence of perfect prediction for all categorical imputation variables. This imputation procedure failed when fitting an ordinal logistic regression model with the message ``convergence not achieved''
% convergence not achieved
% ologit failed to converge on observed data
and no imputations were produced.

\subsection{Simulation study}

A simulation study was conducted to evaluate the performance of a range of strategies for selecting auxiliary variables for inclusion in the univariate imputation models of the MICE procedure. For simplicity, we focused on the setting where there is one incomplete continuous outcome, one completely observed continuous exposure, a set of completely observed continuous auxiliary variables and no confounding. We considered three scenarios of interest: 
\begin{enumerate}
    \item \textsl{Basic:} This scenario was designed to compare the auxiliary variable selection strategies in a scenario where one would not expect the MI algorithm to fail. It is characterised by a large sample size relative to the number of auxiliary variables, moderate missingness mechanism and a moderate (but realistic) proportion of missing data in the outcome. If auxiliary variable selection strategies do not perform well in this scenario, we would not expect them to perform well in general.
    \item \textsl{Extreme:} This scenario was included to compare the auxiliary variable selection strategies in a scenario where including all auxiliary variables in each univariate imputation model would be expected to be problematic. It is characterised by a small sample size relative to the number of auxiliary variables, a strong missingness mechanism and a large proportion of missing data in the outcome. 
    \item \textsl{Realistic:} This scenario was included to compare the auxiliary variable selection strategies in a realistic scenario. The sample size, number of auxiliary variables, missingness mechanism and proportion of missing data were based on what was observed in the motivating LSAC case study.
\end{enumerate}

\subsubsection{Simulation of complete data}

Let $X$ denote a fully observed continuous exposure variable, $Y$ denote a continuous outcome variable with missing values and $\textbf{A} = (A_1, A_2, \dots, A_p)^{\top}$ denote a $p$-vector of potential auxiliary variables. Also let $\boldsymbol{0}_{c}$ denote a $c$-vector of $0$'s. We generated data on $X$, $Y$ and $\textbf{A}$ for $n$ individuals in two steps. In the first step we generated data for $(X, \textbf{A}^{\top})$ by randomly drawing $n$ samples from a multivariate normal distribution with mean $\boldsymbol{0}_{p+1}$ and covariance matrix 
\begin{equation*}
\boldsymbol{\Sigma} = \begin{bmatrix} 1 \ \ & \boldsymbol{0_p}^{\top} \\   \boldsymbol{0_p} \ \ & \boldsymbol{\Sigma}_{A} \end{bmatrix},
\end{equation*}
where $\boldsymbol{\Sigma}_A$ is a $p \times p$ covariance matrix defined for each scenario below. In the second step we generated $Y$ from the linear regression model
\begin{equation}
\label{mod:DGP}
Y = 0.3 X + \boldsymbol{\beta_A}^{\top} \boldsymbol{A} + \varepsilon,
\end{equation}
where $\boldsymbol{\beta_A}$ is a $p$-vector of regression parameters and $\varepsilon \sim N(0, \sigma^2_{\varepsilon})$. 

For the \textsl{Basic} scenario, we set $n = 1000$, $p = 16$ and  $\boldsymbol{\Sigma}_A = \boldsymbol{I}_p$, where $\boldsymbol{I}_p$ is the $p \times p$ identity matrix. For the \textsl{Extreme} scenario, we set $n = 250$, $p = 50$, and $\boldsymbol{\Sigma}_A = \boldsymbol{I}_p$. Since the strength of relationship between an auxiliary variable and a variable with missing data is a key consideration for inclusion in the imputation model,\cite{van2018flexible, graham2012missing, hardt2012auxiliary} we designed these two scenarios such that there was a range of realistic relationships between $Y$ and each auxiliary variable, conditional on $X$ and all other auxiliary variables. 
For the \textsl{Basic} scenario, this was achieved by setting $\boldsymbol{\beta}_A = (\boldsymbol{0}^{\top}_4, 0.1 \boldsymbol{1}^{\top}_4, 0.2 \boldsymbol{1}^{\top}_4, 0.4 \boldsymbol{1}^{\top}_4)^{\top}$, where $\boldsymbol{1}_d$ denotes a $d$-vector of 1's.
For the \textsl{Extreme} scenario, we added 34 additional ``junk'' auxiliary variables in a similar manner to Collins, Schafer and Kam, \cite{collins2001comparison} by setting  $\boldsymbol{\beta}_A = (\boldsymbol{0}_{38}^{\top}, 0.1 \boldsymbol{1}^{\top}_4, 0.2 \boldsymbol{1}^{\top}_4, 0.4 \boldsymbol{1}^{\top}_4)^{\top}$. For ease of interpretation, we specified $\sigma^2_{\varepsilon}$ such that $\boldsymbol{\beta_A}$ was the vector of correlations between $Y$ and $\boldsymbol{A}$. This imposes restrictions on the magnitude of the values that $\boldsymbol{\beta_A}$ can take (equivalent to requiring the covariance matrix of $(Y, X, \boldsymbol{A}^{\top})^{\top}$ to be positive semidefinite). In particular, we could not consider many highly correlated auxiliary variables. However, very large correlations were not observed in the case study (Supporting Information Figure 1).
For the \textsl{Realistic} scenario, we set $n = 4983$, $p = 81$ and $\boldsymbol{\Sigma}_A$ based on the features of the LSAC motivating example.
In this scenario, $\boldsymbol{\beta}_A$ and $\sigma^2_{\varepsilon}$ were specified such that the vector of correlations between $Y$ and $\boldsymbol{A}$ was the same as the estimated vector of correlations between $Y$ and $\boldsymbol{A}$ in the LSAC case study.

We generated 2,000 datasets so that the Monte Carlo standard error for the estimated coverage probability of the 95\% confidence intervals for our estimands (see Section 2.2.3) was less than 0.5\%. Datasets were generated in R. Data for $(X, \boldsymbol{A}^{\top})$ were generated using \texttt{mvrnorm}, which utilises an eigenvalue decomposition of the covariance matrix.

\subsubsection{Imposing missing data}

After generating the complete data, values of $Y$ were set missing with a probability determined by the logistic regression model 
\begin{equation*}
\text{logit} \, P(M_Y) = \gamma_0 + \boldsymbol{\gamma_A}^{\top} \boldsymbol{A},
\end{equation*}
where $M_Y$ denotes the missingness indicator for $Y$ and $(\gamma_0, \boldsymbol{\gamma_A}^{\top}) \in \mathbb{R}^{p+1}$ determines the proportion of missing data, which auxiliary variables are related to $M_Y$ and the strength of missingness mechanism. 

Since we investigated four different strengths of relationship between auxiliary variables and $Y$ in the \textsl{Basic} and \textsl{Extreme} scenarios (via the elements of $\boldsymbol{\beta}_A$), for these scenarios we set $\boldsymbol{\gamma_A}$ such that half of the auxiliary variables with each level of relationship with $Y$ were associated with $M_Y$. For simplicity, the strength of association with missingness was the same for each of these auxiliary variables. For the \textsl{Basic} scenario, values of $Y$ had a 20\% chance of being missing and the odds of missingness increased by 20\% with every one standard deviation increase in each of the auxiliary variables chosen to be associated with missingness (holding the other auxiliary variables fixed). This was achieved by setting the required elements of $\boldsymbol{\gamma_A}$ to either 0 or $\log(1.2)$. The missingness probability and strength of missingness mechanism (i.e., the non-zero values of $\boldsymbol{\gamma_A}$) in this scenario were chosen based on what was observed in the LSAC motivating example (Supporting Information Table 1). 
For the \textsl{Extreme} scenario, values of $Y$ had a 50\% chance of being missing, and the odds of missingness increased by 100\% with every one standard deviation increase in each of the auxiliary variables chosen to be associated with the missingness probability (holding the other auxiliary variables fixed). Since this scenario was included to explore the robustness of the auxiliary variable selection strategies when the full imputation model is likely to encounter difficulties, the missingness mechanism was intentionally chosen to be stronger than what was observed in the case study.
For the \textsl{Realistic} scenario, we specified $\boldsymbol{\gamma}_A$ based on the estimated odds ratio of missingness in \textsl{HRQoL} obtained from univariate logistic regressions of the missingness indicator for \textsl{HRQoL} on the standardised auxiliary variables within the LSAC case study (Supporting Information Table 1). Elements of $\boldsymbol{\gamma}_A$ were set to $\log(1.2)$ if the corresponding odds ratio was greater than 1.15, $\log(0.8)$ if the odds ratio was less than 0.85 and 0 otherwise. 
In Supporting Information Section 2.1 we describe how $\gamma_0$ was calculated to achieve the desired proportions of missingness in $Y$ for each of the simulation scenarios.

\subsubsection{Estimands of interest}

We considered two estimands: the regression coefficient of $X$ on $Y$, i.e., the parameter $\beta_X$ in the following linear regression model 
\begin{equation}
\label{eqn:anamod}
    E(Y \, | \, X) = \beta_0 + \beta_X X,
\end{equation}
and the marginal mean of $Y$, $\mu_Y = E(Y)$. 
The true values of these parameters were taken to be the values used in the data generation.  
That is, $\beta_X = 0.3$ and $\mu_Y = 0$.

The motivation for using MI rather than a complete case analysis may be to reduce bias or increase precision. For the missingness mechanism considered here, the complete case estimate of $\mu_Y$ will be biased, while the complete case estimate of $\beta_X$ will be unbiased.\cite{sterne2009multiple} These points are illustrated graphically Section 2.2 in the Supporting Information, and can also be explained using causal diagrams.\cite{moreno2018canonical} By considering both estimands, we are able to gain insight on the extent to which bias is improved or introduced by the various MI strategies. We are also interested in assessing gains in precision of these estimates from each MI strategy.

\subsubsection{Auxiliary variable selection strategies}
\label{sec:strategies}

We compared the performance of eight data-based strategies for selecting auxiliary variables for an imputation model and two benchmark strategies. 
For each auxiliary variable selection strategy, the variable selection was done once, prior to the imputation step.
The auxiliary variable selection strategies were:
\begin{enumerate}
\item \textsl{Quickpred-pt2:} This approach uses the four-step selection strategy proposed by van Buuren et al.\cite{van1999multiple} For each $i = 1, \dots, p$, $A_i$ is included in the imputation model for $Y$ if the maximum of the absolute correlation between $A_i$ and $Y$ (using all available cases) and the absolute correlation between $A_i$ and $M_Y$ is greater than 0.2. This strategy was implemented using the \texttt{quickpred} function from the R package \texttt{mice}. \cite{buuren2010mice} The relatively low correlation of 0.2 was chosen as a reasonable cut-off for an inclusive strategy.
\item \textsl{Quickpred-pt4:} Similar to \textsl{Quickpred-pt2}, except with the correlation cut-off set to 0.4 instead of 0.2. This cut-off was chosen based on the rule of thumb provided by Graham. \cite{graham2012missing}
\item \textsl{PcAux:} This approach uses principal components of auxiliary variables as predictors in the imputation model instead of the auxiliary variables themselves. \cite{howard2015using}
The number of principal components was chosen such that the principal component scores explained $\geq 40\%$ of the variance in the $A_i$'s, following Howard et al. \cite{howard2015using} 
 This strategy was implemented using \texttt{PcAux} in R.\cite{lang2017pcaux} 
\item \textsl{Forward:} Under this approach, auxiliary variables were chosen using forward selection.  \cite{royston2011multiple} 
An auxiliary variable was added to the imputation model for $Y$ at each step of the forward selection algorithm if the $p$-value from a Wald test on the corresponding regression coefficient was less than 0.05. 
This strategy was implemented in Stata using the \texttt{stepwise} option in \texttt{ice}.
\item \textsl{Forward-sw:} This approach is similar to \textsl{Forward}, except that at each step auxiliary variables could either be added to the imputation model as per above, or removed if the $p$-value from the Wald test dropped below 0.05. It was also implemented using \texttt{ice} in \texttt{Stata}.
\item \textsl{Forward-FMI:} This approach uses forward selection based on the FMI in the mean of $Y$ (hereafter ``FMI'') as proposed by Andridge and Thompson. \cite{andridge2015using}
Here we provide a brief description of the theory and implementation of this strategy in the current study. For further details and extensions under different assumptions we refer the reader to the work of Andridge and Thompson,\cite{andridge2015using} and Little.\cite{ little1994class}
Let $A^*$ denote a ``proxy variable'' which is created as the predicted values from a linear regression of $Y$ on a subset of $\boldsymbol{A}$. Assume that $M_Y$ follows a Bernoulli distribution and that the distribution of $(Y, A^*)$, given $M_Y$, is bivariate normal with distinct means and variances for respondents ($M_Y = 0$) and nonrespondents ($M_Y = 1$). Also assume that the probability that $Y$ is missing depends only on $A^*$. Under these assumptions, the maximum likelihood estimates of $\mu_Y$ and the FMI can be expressed as functions of the observed data.
The forward selection procedure was implemented as follows.
The procedure was initialised by estimating the FMI for each auxiliary variable in turn and selecting for inclusion in the imputation model the auxiliary variable associated with the smallest FMI.
In the next step, $p - 1$ pairs of auxiliary variables (with each pair consisting of the auxiliary variable chosen in the previous step and one of the remaining $p - 1$ auxiliary variables) were used to create $p - 1$ proxy variables. Estimates of the FMI were obtained for each proxy variable, with the auxiliary variable from the pair resulting in the smallest estimated FMI added to the imputation model.
 The forward selection procedure continued in this manner until the reduction in the FMI was less than a pre-specified percentage of the missing data in $Y$. The justification for the use of this stopping rule and further details are provided in Section 2.4 of the Supporting Information.
\item \textsl{Tests:} Under this approach, auxiliary variables were selected for inclusion in the imputation model for $Y$ if the $p$-value obtained from a t-test for a given auxiliary variable, carried out using $M_Y$ to define the two groups, was less than 0.05. 
\item \textsl{LASSO:} In this approach, a model for $Y$ that included $X$ and all auxiliary variables was fitted using the LASSO with 10-fold cross-validation.\cite{friedman2010} The regularisation penalty (usually denoted by $\lambda$) was chosen to be the value that gave the most regularised model such that the cross-validation error was within one standard error of the minimum.\cite{james2013introduction} %p236
The auxiliary variables that were included in the fitted model were included in the imputation model for $Y$, i.e., the LASSO was used to select variables for the imputation model but not to estimate parameters of the imputation model. 
This is similar to the ``indirect use
of regularised regression'' approach described elsewhere.\cite{zhao2016multiple} 
This strategy was implemented using \texttt{glmnet}.
\end{enumerate}
The benchmark strategies were:
\begin{enumerate}
\item \textsl{CCA:} A complete case analysis. The subset of individuals included in the analysis was restricted to those with data for $Y$. 
\item \textsl{Full:} All auxiliary variables were included in the imputation model for $Y$. 
\end{enumerate}

To be useful in practice, an auxiliary variable selection strategy should perform better than the \textsl{CCA} strategy by either reducing bias and/or increasing precision (see Discussion section for further comments). Ideally the auxiliary variable selection strategy would also perform no worse than the \textsl{Full} strategy if the imputation algorithm for this strategy does not fail, and successfully produce imputations in the case that it does fail. 

All imputation models included $X$ and $m = 50$ imputations were performed, which was chosen to be equal to the percentage of missing data in $Y$ in the \textsl{Extreme} scenario.\cite{royston2011multiple} 
MI estimates were obtained using Rubin's rules.\cite{rubin2004multiple} Unless stated otherwise, MI was implemented using \texttt{mice} with linear regression in R.\cite{buuren2010mice}

\subsubsection{Performance measures}

The performance of the 10 analysis strategies (eight auxiliary variable selection strategies and two benchmark strategies) for estimating $\beta_X$ and $\mu_Y$ was compared for each of the three simulation scenarios using bias, empirical standard error (SE), average model SE, and coverage probability of 95\% confidence intervals. We used the definitions of these performance measures, as well as the formula for Monte Carlo SE estimates, that are given in Table 6 of Morris et al. \cite{morris2019using} %Page 39 
We also report standardised bias, defined as (bias / empirical SE) $\times$ 100, relative bias for $\beta_X$, defined as (bias / true parameter value) $\times$ 100 (this can not be calculated for $\mu_Y = 0$), the relative error in the model SE, defined as (average model SE / empirical SE - 1) $\times$ 100, and
the convergence rate, defined as the proportion of datasets for which MI estimates were successfully obtained.

\subsection{Application of strategies to LSAC example}

To illustrate their performance in practice, we applied each of the auxiliary variable selection strategies considered in the simulation study to the LSAC case study described in Section \ref{sec:LSAC}. We also carried out the benchmark strategies \textsl{CCA} and \textsl{Full} (in R).
We used MICE with the same univariate regression models that were specified in Section \ref{sec:LSAC}.
All imputation models included all analysis model variables (to ensure congeniality), with auxiliary variables chosen according to the analysis strategy. A total of 20 imputations were performed, with 10 iterations for each imputation.
In the next paragraph, we outline how each auxiliary variable selection strategy was implemented to accommodate the additional complexities of multivariate missingness and different types of variables.

For the \textsl{Quickpred-pt2} and \textsl{Quickpred-pt4} strategies, the \texttt{quickpred} function was used to select the auxiliary variables in the imputation model for each of the incomplete variables using the correlation cut-offs of 0.2 and 0.4, respectively.
The \textsl{PcAux} strategy required imputation of the missing values in the auxiliary variables prior to obtaining principal components. This was done using the \texttt{PcAux} package,\cite{lang2017pcaux} which implements a single imputation using PMM within the MICE approach. The first eight principal components of the auxiliary variables, which explained approximately 44\% of the variance in these variables, were included in the imputation models for the remaining five incomplete variables in the analysis model.
For the \textsl{Forward} and \textsl{Forward-sw} strategies, the stepwise algorithm was used to select the auxiliary variables in the imputation model for each of the incomplete variables. Dummy variables for the ordinal PedsQL items were grouped during the variable selection step to ensure they were either all included in, or excluded from, the imputation model. 
The \textsl{Forward-FMI} method does not currently handle missing values in auxiliary variables or highly negatively skewed ordinal variables.\cite{andridge2015using} Rather than extend this methodology (which is beyond the scope of this study) we deemed this method not suitable to use in this case study.
For the \textsl{Tests} strategy, t-tests (for continuous variables) or chi-squared tests (for categorical variables) were used to select the auxiliary variables in the imputation model for each of the incomplete variables, with the two groups to be compared determined by the missingness indicator of the incomplete variable. An auxiliary variable was selected for inclusion in the imputation model for the incomplete variable if the $p$-value obtained from the hypothesis test was less than 0.05 and a sufficient number of observations were available in each group ($\geq 20$ for the t-test and all expected counts $\geq 5$ for the chi-squared test).
For the \textsl{LASSO} strategy, we wanted to implement the grouped LASSO to ensure dummy variables for PedsQL items were either all included or excluded from the imputation model (c.f. \textsl{Forward} and \textsl{Forward-sw}). However, it is not clear how to apply the grouped LASSO with ordinal outcomes. Furthermore, the LASSO requires predictors (here the analysis model variables and the candidate auxiliary variables) to be complete. Rather than imputing the auxiliary variables, we proceeded as follows. A model for \textsl{HRQoL} that included all variables in the analysis model and all candidate auxiliary variables was fitted to the complete cases using the grouped LASSO with 10-fold cross-validation (implemented using the \texttt{gglasso} package in R). 
The regularisation penalty was chosen to be the value that gave the most regularised model such that the cross-validation error was within one standard error of the minimum.
Each auxiliary variable selected using the LASSO was included in the imputation model for each of the incomplete variables in the analysis model, as well as the imputation model for all other incomplete auxiliary variables selected by the LASSO.
Auxiliary variables not selected by the LASSO were excluded from the imputation process altogether.

\section{Results}

\subsection{Simulation study}

\textbf{Mean of $Y$}

\begin{figure}
    \centering
    \includegraphics[width = 0.9\textwidth]{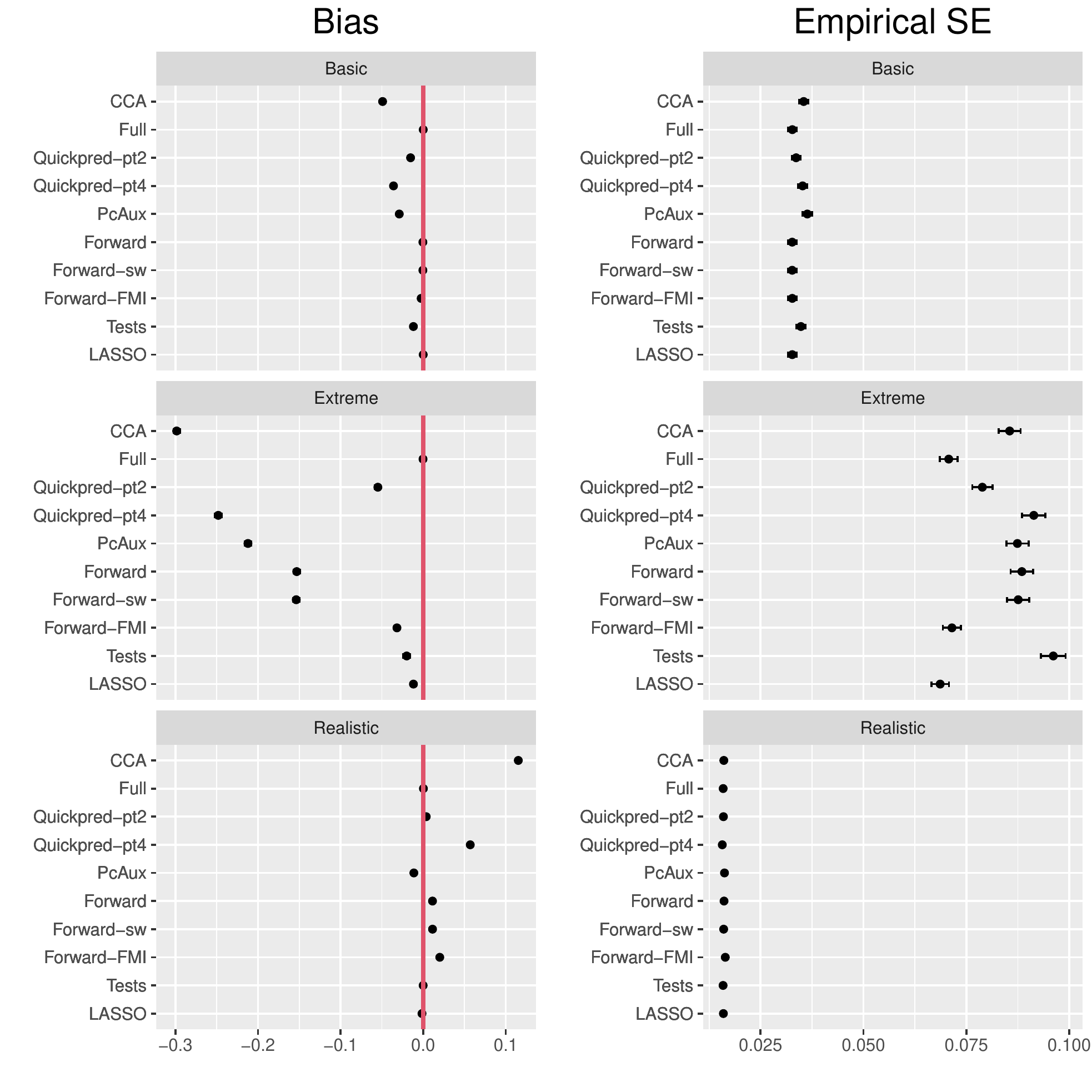}
    \caption{Simulation estimates of bias and empirical SE for $\mu_Y$ (true value = 0) from each analysis strategy and for each scenario. Estimates are presented with 95\% Monte Carlo confidence intervals to quantify simulation uncertainty.}
    \label{fig:meany}
\end{figure}

Figure \ref{fig:meany} presents simulation estimates of bias and empirical SE for $\mu_Y$ (true value equal to 0), with 95\% Monte Carlo confidence intervals, for each analysis strategy and scenario. Full results are presented in Table 2 in the Supporting Information. There were no problems with convergence of any of the analysis approaches, with estimates of $\mu_Y$ (and $\beta_X$) produced on all simulation runs. 
% Bias
As expected, estimating $\mu_Y$ using complete cases led to the largest bias, while using MI with all auxiliary variables in the imputation model was unbiased across all scenarios. Within each scenario, all auxiliary variable selection strategies led to smaller absolute bias than the complete case analysis, with absolute bias from the \textsl{Quickpred-pt2}, \textsl{Forward-FMI}, \textsl{Tests} and \textsl{LASSO} strategies less than one third of the absolute bias from the complete case analysis in each scenario. The smallest standardised bias was observed for the \textsl{Full} and \textsl{LASSO} strategies (less than 30\% in each scenario, an amount which is considered not to be problematic by Collins, Schafer and Kam).\cite{collins2001comparison}
% Empirical SE
MI with all auxiliary variables in the imputation model led to lower empirical SEs than the complete case analysis across each scenario, as did the \textsl{Quickpred-pt2} and \textsl{LASSO} strategies. The remaining MI strategies had larger empirical SEs compared to the complete case analysis in at least one of the scenarios (with \textsl{PcAux} producing larger empirical SEs in all three scenarios).
% Average model SE
All strategies in the \textsl{Basic} and \textsl{Realistic} scenarios produced average model SEs that underestimated the empirical SEs, although the absolute value of the relative error in the model SE did not exceed 10\%. In the \textsl{Extreme} scenario, relative error in the model SE was underestimated by more than 10\% for the \textsl{Quickpred-pt4}, \textsl{Forward} and \textsl{Forward-sw} strategies (-16\%, -20\% and -19\%, respectively).
% Coverage
Estimating $\mu_Y$ using complete cases led to the lowest coverage probability in each scenario (ranging from 0\% in the \textsl{Realistic} scenario to 70\% in the \textsl{Basic} scenario), followed by the \textsl{Quickpred-pt4} strategy (ranging from 4\% in the \textsl{Realistic} scenario to 80\% in the \textsl{Basic} scenario). In comparison, the \textsl{Full}, \textsl{Tests} and \textsl{LASSO} strategies had reasonable coverage probability in each scenario (between 93--96\%). Note that comparing coverage probabilities across scenarios is difficult due to different sample sizes.

\newpage

\noindent \textbf{Regression coefficient of $X$}

\begin{figure}
    \centering
    \includegraphics[width = 0.9\textwidth]{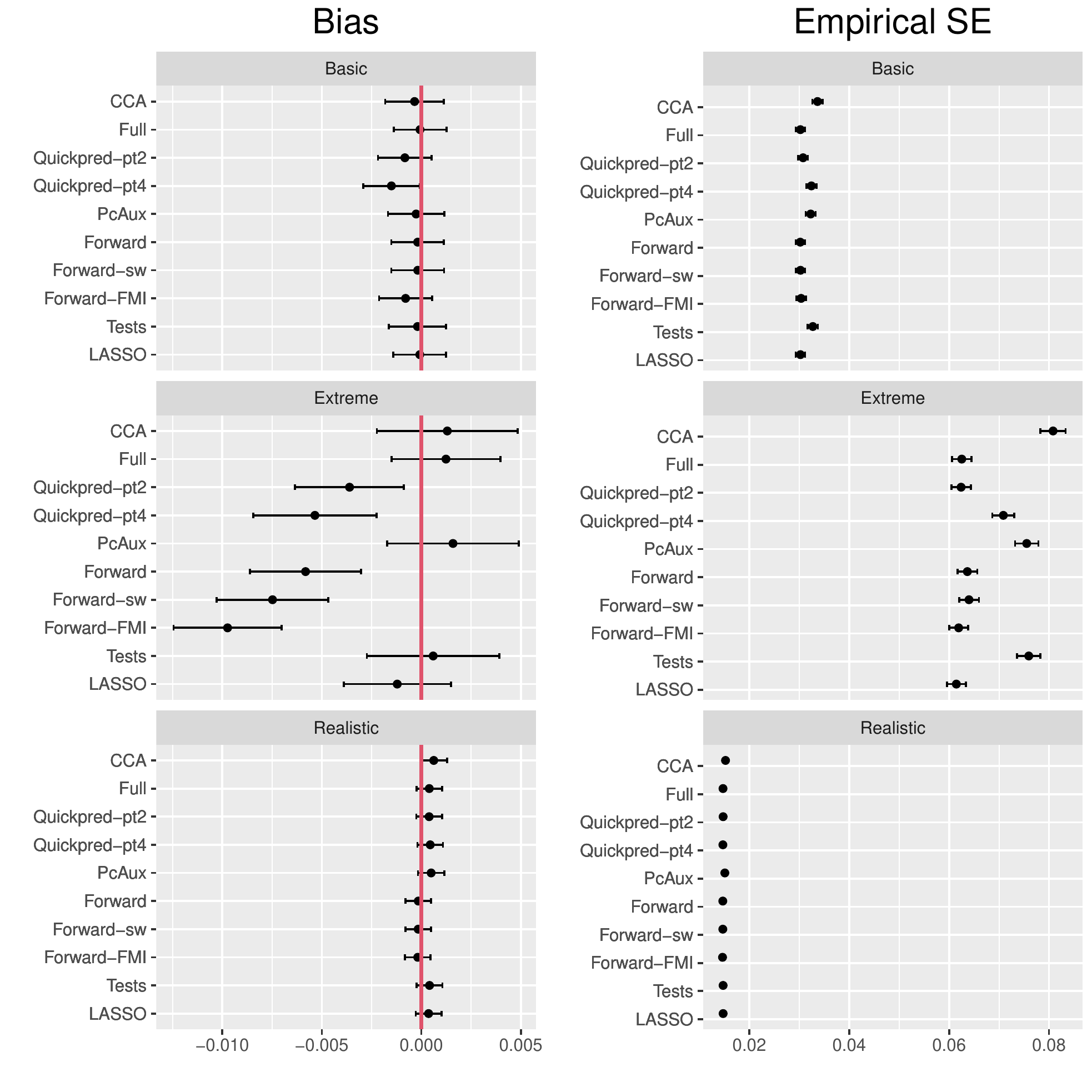}
    \caption{Simulation estimates of bias and empirical standard error (SE) for $\beta_X$ (true value = 0.3) from each analysis strategy and for each scenario. Estimates are presented with 95\% Monte Carlo confidence intervals to quantify simulation uncertainty.}
    \label{fig:betax}
\end{figure}

Figure \ref{fig:betax} presents simulation estimates of bias and empirical SE for $\beta_X$ (true value equal to 0.3), with 95\% Monte Carlo confidence intervals, for each analysis strategy and scenario. Full results are presented in Table 2 in the Supporting Information. 
% Bias 
As expected, both the complete case analysis and MI with all auxiliary variables resulted in negligible bias in all scenarios. All other MI strategies also resulted in negligible bias, with the largest absolute biases (which were obtained in the \textsl{Extreme} scenario) amounting to less than 4\% of the true parameter value and all standardised biases less than 16\%.
% Empirical SE
All MI strategies produced smaller empirical SEs than the complete case analysis, with gains (absolute differences) in precision of up to 0.0035 in the \textsl{Basic} scenario, 0.019 in the \textsl{Extreme} scenario and 0.0006 in the \textsl{Realistic} scenario.
% Average model SE
The model SE was a reasonable estimate of the empirical SE for all strategies and scenarios. The relative error in the average model SE did not exceed 2\% across analysis strategies in the \textsl{Basic} scenario, 7\% in the \textsl{Extreme} scenario or 4\% in the \textsl{Realistic} scenario.
% Coverage
All coverage probabilities were between 94\% and 96\%.

\medskip

\noindent \textbf{Selected auxiliary variables}

Table \ref{tab:1} provides further information on the auxiliary variables selected by the analysis strategies on each simulation run. This table does not include \textsl{PcAux} since this strategy included auxiliary variable information via principal component scores. 
In the \textsl{Basic} and \textsl{Extreme} scenarios, all strategies except \textsl{Tests} and \textsl{Quickpred-pt4} included auxiliary variables that had a correlation $\geq 0.4$ with $Y$ and were not related to missingness on most ($>90$\%) of the simulation runs. The \textsl{Tests} strategy selected these variables relatively rarely. The \textsl{LASSO} strategy (1) was the most inclusive strategy overall (determined by the total average number of auxiliary variables included) and (2) included the largest number of auxiliary variables that were neither correlated with $Y$ or related to $M_Y$, although this number was still relatively low compared to the number of auxiliary variables in this category. In the \textsl{Realistic} scenario, the \textsl{Tests} strategy was the most inclusive overall, followed by \textsl{Quickpred-pt2} and then \textsl{LASSO}.

\medskip

% latex table generated in R 4.1.0 by xtable 1.8-4 package
% Tue Jan 04 12:44:02 2022
\begin{table}[ht]
\centering
\resizebox{0.98\columnwidth}{!}{\begin{tabular}{llllllllll}
  \hline
  \multirow{2}{*}{\textbf{Strategy}} & \multicolumn{8}{c}{\textbf{Correlation between $A_i$ and $Y$; $A_i$ associated with $M_Y$}} & \multirow{2}{*}{\textbf{Total}} \\
& 0; yes   & 0; no   & 0.1; yes   & 0.1; no   & 0.2; yes   & 0.2; no   & 0.4; yes  & 0.4; no & \\ \hline
\textbf{Basic, n} & \textbf{2} & \textbf{2} & \textbf{2} & \textbf{2} & \textbf{2} & \textbf{2} & \textbf{2} & \textbf{2} & \textbf{16} \\ 
  Quickpred-pt2 & 0 (0) & 0 (0) & 0 (0.1) & 0 (0.3) & 0.9 (43.6) & 1 (51.5) & 2 (100) & 2 (100) & 5.9 (36.9) \\ 
  Quickpred-pt4 & 0 (0) & 0 (0) & 0 (0) & 0 (0) & 0 (0) & 0 (0) & 0.9 (46.2) & 1.1 (53.8) & 2 (12.5) \\ 
  Forward & 0.1 (5.7) & 0.1 (4.6) & 1.9 (95.4) & 1.9 (97.2) & 2 (100) & 2 (100) & 2 (100) & 2 (100) & 12.1 (75.4) \\ 
  Forward-sw & 0.1 (5.7) & 0.1 (5.1) & 1.9 (95.6) & 1.9 (97.3) & 2 (100) & 2 (100) & 2 (100) & 2 (100) & 12.1 (75.5) \\ 
  Forward-FMI & 0 (0) & 0 (0) & 1.3 (63.2) & 1.3 (65.3) & 2 (100) & 2 (100) & 2 (100) & 2 (100) & 10.6 (66.1) \\ 
  Tests & 1.2 (59.7) & 0.1 (5.5) & 1.2 (59.5) & 0.1 (4.9) & 1.2 (59.8) & 0.1 (4.9) & 1.2 (59.6) & 0.1 (5.4) & 5.2 (32.4) \\ 
  LASSO & 0.2 (11.3) & 0.2 (11.3) & 2 (100) & 2 (100) & 2 (100) & 2 (100) & 2 (100) & 2 (100) & 12.5 (77.8) \\ 
   \\ \textbf{Extreme, n} & \textbf{2} & \textbf{36} & \textbf{2} & \textbf{2} & \textbf{2} & \textbf{2} & \textbf{2} & \textbf{2} & \textbf{50} \\ 
  Quickpred-pt2 & 1.2 (61.2) & 1 (2.7) & 1.2 (60) & 0.3 (14.3) & 1.4 (70.2) & 1.1 (54.1) & 2 (98.9) & 2 (99.6) & 10.2 (20.3) \\ 
  Quickpred-pt4 & 0 (0.1) & 0 (0) & 0 (0) & 0 (0) & 0 (0.1) & 0 (0.9) & 0.6 (30.2) & 1.2 (60.2) & 1.8 (3.7) \\ 
  Forward & 0.2 (8.6) & 3.7 (10.3) & 0.1 (5.7) & 0.3 (15) & 0.5 (22.6) & 0.9 (43.3) & 1.7 (83.5) & 1.9 (92.8) & 9.1 (18.3) \\ 
  Forward-sw & 0.2 (9.4) & 3.7 (10.4) & 0.1 (6.1) & 0.3 (15.2) & 0.5 (22.7) & 0.9 (42.7) & 1.7 (82.8) & 1.9 (93.4) & 9.2 (18.4) \\ 
  Forward-FMI & 0.1 (7.5) & 1.8 (5) & 0.9 (42.8) & 1.3 (64.2) & 1.9 (96.7) & 2 (99.2) & 2 (100) & 2 (100) & 12 (24) \\ 
  Tests & 1.9 (93.2) & 1.8 (5) & 1.8 (92.5) & 0.1 (5) & 1.9 (93) & 0.1 (5) & 1.9 (92.8) & 0.1 (4.3) & 9.5 (19) \\ 
  LASSO & 0.5 (26.4) & 7.5 (20.8) & 1.9 (93.3) & 2 (97.7) & 2 (100) & 2 (100) & 2 (100) & 2 (100) & 19.8 (39.7) \\ 
   \\   & \multicolumn{8}{c}{\textbf{Interval containing correlation between $A_i$ and $Y$; $A_i$ associated with $M_Y$}} & \\  & \multicolumn{1}{l}{{[}0, 0.1); yes} & \multicolumn{1}{l}{{[}0, 0.1); no} & \multicolumn{1}{l}{{[}0.1, 0.2); yes} & \multicolumn{1}{l}{{[}0.1, 0.2); no} & {[}0.2, 0.4); yes & {[}0.2, 0.4); no & {[}0.4, 1{]}; yes & {[}0.4, 1{]}; no & \\ \cline{2-9}\textbf{Realistic, n} & \textbf{3} & \textbf{0} & \textbf{11} & \textbf{23} & \textbf{13} & \textbf{28} & \textbf{3} & \textbf{0} & \textbf{81} \\ 
  Quickpred-pt2 & 0.9 (29.9) & - & 4.2 (38.5) & 1.5 (6.7) & 12.5 (95.9) & 24.2 (86.4) & 3 (100) & - & 46.3 (57.2) \\ 
  Quickpred-pt4 & 0 (0) & - & 0 (0) & 0 (0) & 0 (0) & 0 (0) & 2.1 (70.2) & - & 2.1 (2.6) \\ 
  Forward & 0.3 (10.3) & - & 3 (27.2) & 5.3 (23.3) & 4.1 (31.8) & 9.1 (32.6) & 1.3 (43.8) & - & 23.2 (28.7) \\ 
  Forward-sw & 0.3 (10.8) & - & 3.1 (27.8) & 5.3 (23) & 4.1 (31.2) & 9 (32.1) & 1.4 (45.4) & - & 23.1 (28.5) \\ 
  Forward-FMI & 0 (1.6) & - & 1.1 (10.2) & 3.6 (15.6) & 2.9 (22.4) & 7.8 (27.9) & 1.4 (47.9) & - & 16.9 (20.9) \\ 
  Tests & 3 (100) & - & 10.1 (91.8) & 22.9 (99.3) & 12.9 (99.5) & 28 (100) & 3 (100) & - & 79.9 (98.6) \\ 
  LASSO & 0.4 (13.8) & - & 6.1 (55.4) & 8.7 (37.9) & 8.7 (66.7) & 14.1 (50.4) & 1.2 (39.4) & - & 39.2 (48.4) \\ 
   \hline
\end{tabular}}
\caption{Average number (\%) of times auxiliary variables were selected across simulation runs for relevant analysis strategies. Auxiliary variables are grouped by correlation with $Y$ and association with the indicator for missingness in $Y$ ($M_Y$). The scenario and corresponding number of auxiliary variables in each group is provided in bold.} 
\label{tab:1}
\end{table}

\subsection{LSAC example}

\begin{figure}
    \centering
    \includegraphics[width = 0.9\textwidth]{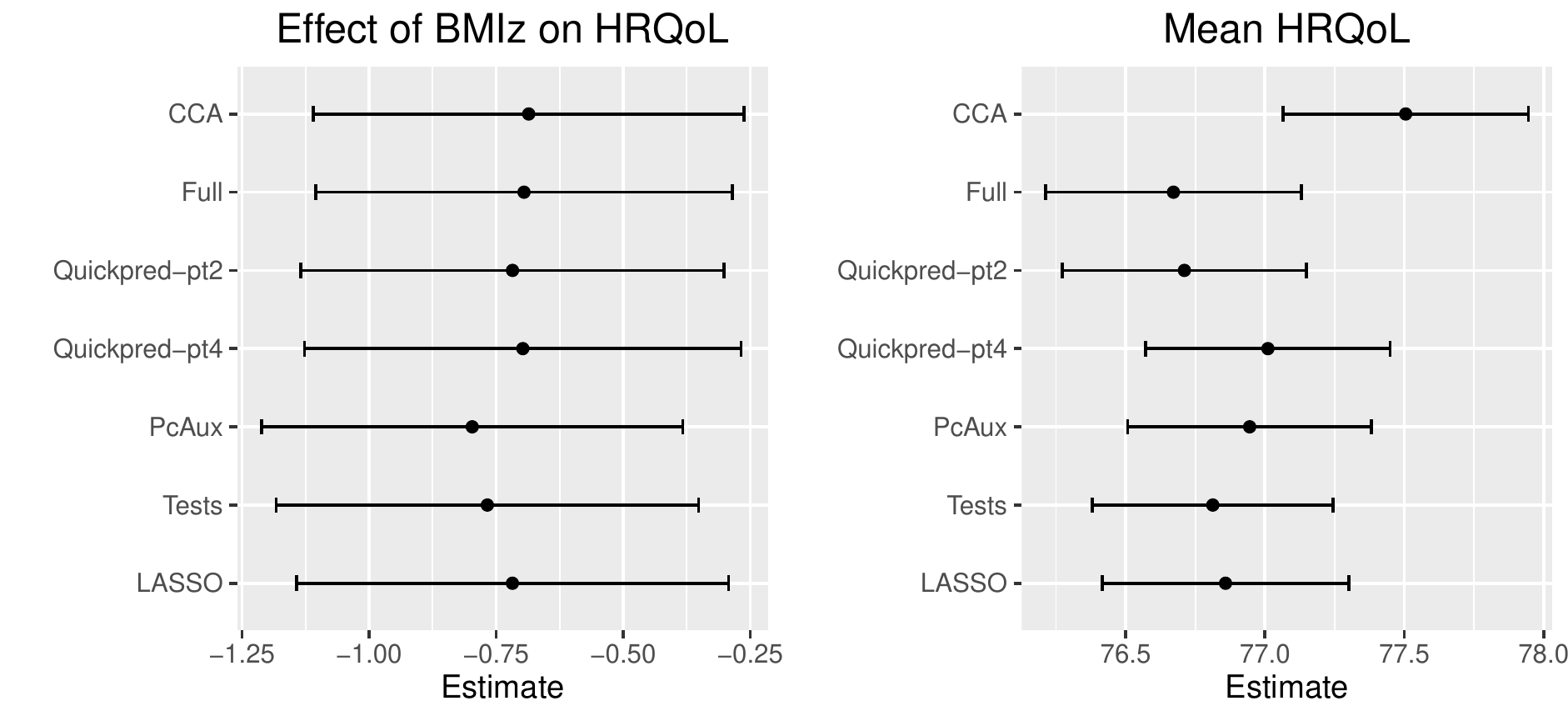}
    \caption{Estimates of the effect of BMIz on HRQoL (left panel) and mean HRQoL (right panel), with 95\% confidence intervals, from seven different analysis strategies.}
    \label{fig:LSAC}
\end{figure}

Figure \ref{fig:LSAC} presents estimates, with 95\% confidence intervals, for the effect of \textsl{BMIz} on \textsl{HRQoL} ($\beta_1$ in model \eqref{eqn:anamod}) and the mean \textsl{HRQoL} for the LSAC case study, using seven of the analysis strategies. Estimates from \textsl{Forward} and \textsl{Forward-sw} are not presented in this figure as the MI algorithm for these strategies (which were implemented in Stata) failed with the message ``convergence not achieved'' during the auxiliary variable selection step for one of the PedsQL items and no estimates were produced. Although the R strategies produced estimates, problems were reported (via the list of logged events reported by \texttt{mice}) in the \textsl{Full} and \textsl{Tests} strategies. No problems were reported for the other MI strategies.
All analysis strategies led to similar estimates of $\beta_X$ (ranging from -0.73 to -0.69) and therefore similar conclusions. That is, a one-unit increase in BMI z-score in Australian children aged 4 to 5 years is associated with a decrease in HRQoL at age 10 to 11 years of around 0.7 units. Estimates of the mean HRQoL ranged from 76.7 to 77.5, with estimates produced by all MI strategies noticeably lower than estimates from the complete case analysis.
In Table 3 in the Supporting Information we report the variables selected for inclusion in the imputation model for \textsl{HRQoL} for the MI strategies that produced estimates. Of the 85 candidate auxiliary variables considered, 61 were selected for inclusion in the imputation model for \textsl{HRQoL} by the \textsl{Quickpred-pt2} strategy, 3 by the \textsl{Quickpred-pt4} strategy, 50 by the \textsl{Tests} strategy and 17 by the \textsl{LASSO} strategy.

\section{Discussion}

% Summarise results
Although data-driven auxiliary variable selection is commonly used in the application of MI, to our knowledge this study is the first comprehensive investigation of such strategies.
We compared the performance of eight data-based strategies for selecting auxiliary variables for inclusion in the imputation procedure when MI was used to estimate the mean outcome and an exposure-outcome association when the outcome had missing values, under three scenarios of interest. We also carried out the analysis using complete cases and MI with all auxiliary variables in the imputation model. MI with the full imputation model performed well across all scenarios considered, regardless of whether the target of analysis was the exposure-outcome association or the mean outcome, and did not fail to converge despite the \textsl{Extreme} scenario. 
All auxiliary variable selection strategies led to negligible bias and smaller empirical standard errors than the complete case analysis when estimating the exposure-outcome association in the simulation study.
When estimating the mean, the best performing strategy (based on bias and coverage probability across all scenarios) was the \textsl{LASSO}. This strategy had the advantage of incorporating all auxiliary variable information when selecting variables for the imputation model (rather than information from multiple pairwise comparisons). However, it was not clear how to best extend this approach to the case study where there were the additional complexities of missing data in auxiliary variables and highly skewed ordinal variables. 

% Our work in the context of existing auxiliary variable selection literature
This study adds to the limited body of research on auxiliary variable selection for MI. 
Early work in this area illustrated that auxiliary variables can improve the performance of MI and led to the recommendation of an inclusive strategy for variable selection.\cite{collins2001comparison} 
% Link to theoretical arguments and other arguments for recommending inclusive strategy?
However, the study by Collins et al.\ considered just two auxiliary variables that were correlated with either the incomplete variable or the missingness of the incomplete variable. In large-scale longitudinal studies there are often many potential auxiliary variables to choose from. The results of the current study provide support for the inclusive auxiliary variable strategy when there are a larger number of auxiliary variables and a mix of realistic relationships between these variables and an incomplete outcome. This is illustrated, for example, by the \textsl{Full} strategy performing better than the \textsl{Quickpred-pt2} strategy, which in turn performs better than the \textsl{Quickpred-pt4} strategy in terms of bias across scenarios.
The quickpred strategies are attractive in practice due to their simplicity, the ease at which they extend to more complicated problems (such as the LSAC case study) and their implementation in statistical software.
In contrast, although the \textsl{Forward-FMI} strategy performed reasonably in this study, 
there is currently no software freely available to apply it in practice and the strategy does not easily extend to complicated problems.
Our results also show that if one employs the \textsl{Tests} strategy, auxiliary variables that are strongly associated with the incomplete variables may be omitted from the imputation model, which results in larger standard errors (relative to other MI strategies).

% Other strategies for getting the imputation model to run
Although appealing, employing an inclusive strategy may be problematic when there are many auxiliary variables available. We encountered convergence problems when analysing the case study with Stata using the \textsl{Full}, \textsl{Forward} and \textsl{Forward-sw} strategies. Results were obtained when the \textsl{Full} strategy was implemented in R, but with a number of logged events reported on each iteration of the imputation algorithm.
Others have also noted convergence problems when using MICE in Stata,\cite{nguyen2021practical, simons2015multiple} 
and one simulation study using MICE in both Stata and R found that, although R did not encounter the same convergence problems as Stata, the estimates obtained in R were less accurate than estimates from a MVNI-based approach.\cite{de2021multiple}
The lack of convergence problems in our simulation study is likely due to the simulated datasets containing only continuous variables (and only one incomplete variable), with no very high correlations between these variables.
Instead of reducing or modifying the set of auxiliary variables, one could try several alternative approaches to produce imputations when the MI algorithm fails. These include changing the imputation model form, e.g., instead of ordinal logistic regression to impute ordinal variables, one could try linear regression or PMM; changing the imputation method, e.g. for missing data in more than one variable, MVNI could be used instead of multiple imputation by chained equations \cite{lee2010multiple}; or changing the level at which variables are imputed (applicable for derived variables or scale scores). \cite{nguyen2021practical, mainzer2021comparison}
However, these approaches also have limitations.
Perhaps the most effective approach for overcoming challenges with the MI algorithm would be to identify and then remove from the imputation model the problematic variables. Although, if these appear in the substantive analysis then uncongeniality of the imputation model and substantive model becomes a concern.
We did not consider such alternative approaches in this study.

% Auxiliary variable selection
Many auxiliary variable selection strategies have been proposed and implemented in statistical software, but there has been little empirical evidence to evaluate these strategies even though data-based variable selection strategies are known to have limitations in other contexts.\cite{kabaila2009coverage, greenland1989modeling}
One challenge in evaluating such strategies is designing a study to include a mix of realistic correlations between auxiliary variables and variables in the analysis model.
A study by Hardt et al. considered two values of the correlations between all auxiliary variables and other variables in the study: low (0.1) and moderate (0.5).\cite{hardt2012auxiliary}
Another study by Howard et al. took a similar approach, varying correlation values from 0.1 to 0.6.\cite{howard2015using}
In practice, it is highly unlikely for all auxiliary variables to have the same relationship with the variable with missing values. Our simulation study was designed to assess the performance of strategies for selecting auxiliary variables when there were a mix of relationships between auxiliary variables and (i) the incomplete variable, and (ii) the missingness indicator of the incomplete variable.
In contrast to our study, both Collins et al. and Hardt et al. showed that including many auxiliary variables in the imputation model can lead to bias in MI estimates.\cite{collins2001comparison, hardt2012auxiliary} 
We did not observe bias from our \textsl{Full} analysis strategy in any scenario considered.
However, our study was not set up to address limits on the number of auxiliary variables that should be included. 
%Another way to prevent problems caused by including too many auxiliary variables in the imputation procedure is to reduce the number of potential auxiliary variables by identifying a subset of potential auxiliary variables based on substantive knowledge. This approach was taken in our case study.
The results of the current study also reiterate that the decision of whether to apply MI should include careful consideration of whether the estimand can be estimated without bias from the available data.
Causal diagrams that include missingness indicators are a useful tool to aid this decision.\cite{moreno2018canonical}
If the objective of a study is to estimate an exposure-outcome association when there is missing data in the outcome and no useful auxiliary variables are available, a complete case analysis may be sufficient.
In this paper we considered an auxiliary variable strategy to perform better than the complete case analysis if it reduced bias and/or increased precision compared to the complete case analysis. However, we note that there are cases in which MI will (appropriately) result in larger standard errors than a complete case analysis as the uncertainty in the missing values is taken into account. 

% Tuning parameters
There are many variations on the ways in which the auxiliary variable selection strategies may be implemented that were not considered in this paper. For the quickpred strategies, one could consider adjusting the correlation cut-off such that a specified number of auxiliary variables with the highest correlations are chosen or exclude auxiliary variables with too many missing values.\cite{van2018flexible, van1999multiple} For the \textsl{PcAux} strategy, the number of principal components to include in the imputation model was chosen such that the principal component scores explained $\geq 40\%$ of the variance in the auxiliary variables. One may consider a different proportion of variance explained or employ other methods for specifying the number of principal components to be used. The \texttt{pcaux} function provides an option (\texttt{nComps = Inf}) to use the smallest number of component scores such that adding one more component score does not make a discernable difference in the amount of variance explained. Also, the $p$-value threshold may be changed in the \textsl{Forward}, \textsl{Forward-sw} and \textsl{Tests} strategies to include different subsets of auxiliary variables. Finally, we used the LASSO for variable selection only, since the aim of the study was to compare methods used to select auxiliary variables for MI. However, there are alternative ways to incorporate the LASSO with MI that may perform better.\cite{zhao2016multiple}

% Strengths and limitations
This study provides a natural starting point for evaluating a range of auxiliary variable selection strategies for MI. 
However, several limitations of this study should be noted.
Firstly, all variables in our simulation study were continuous and normally distributed. When there are a large number of categorical auxiliary variables (such as in the case study), perfect prediction can cause problems with MI. Others have considered how to do MI in this scenario.\cite{nguyen2021practical, mainzer2021comparison}
For simplicity, we considered the exposure and auxiliary variables to be independent which may not be realistic in practice. 
An alternative approach may be to generate data using causal diagrams and a series of regression models.
We also acknowledge that, (although not the case in the current study) including auxiliary variables may introduce bias into the MI estimates because of underlying relationships between variables.\cite{thoemmes2014cautious} 
The strength of missing data mechanism was the same for each auxiliary variable, there was only one incomplete variable and there was no confounding. These simplifying assumptions are unlikely to hold in practice.
Finally, we chose to focus on only three scenarios. By doing so, we may have overlooked a range of scenarios under which the analysis strategies perform poorly. 
However, this study can be used to inform further research in this area.

% Conclusion
In conclusion, this study evaluated a range of strategies for selecting auxiliary variables for MI models, with the aim of providing advice for researchers faced with this problem. 
MI using all auxiliary variables in the imputation model performed well in all scenarios considered, providing support for adopting an inclusive auxiliary variable strategy where possible. 
Auxiliary variable selection using the LASSO was the best performing auxiliary variable selection strategy overall and is a promising alternative when the full model fails.
Quick data-based selection of auxiliary variable, as implemented in \texttt{quickpred}, also performed reasonably when used with a low cut-off and was straightforward to apply to the LSAC case study. 
However we note that further work is needed to evaluate these strategies in a range of different scenarios.

\section*{Acknowledgments}
This paper uses unit record data from Growing Up in Australia: the Longitudinal Study of Australian Children (LSAC). LSAC is conducted by the Australian Government Department of Social Services (DSS). The findings and views reported in this paper, however, are those of the authors and should not be attributed to the Australian Government, DSS, or any of DSS’ contractors or partners. DOI: 10.26193/F2YRL5. This work was supported by an Australian National Health and Medical Research Council (NHMRC) Career Development Fellowship (CDF) Level 2 Grant (grant 1127984) and a NHMRC Project Grant (grant 1166023). Ian White was supported by the Medical Research Council (grant number MC\_UU\_00004/07). Research at the Murdoch Children’s Research Institute is supported by the Victorian Government’s Operational Infrastructure Support Program.

\subsection*{Author contributions}

Rheanna Mainzer: Conceptualization, Programming, Formal analysis, Visualization, Writing - Original Draft.
Katherine Lee:  Conceptualization, Supervision.
All authors provided input into the design of the study, feedback on the manuscript, final approval of the version to be published and agreement to be accountable for all aspects of the work in ensuring that questions related to the accuracy or integrity of any part of the work are appropriately investigated and resolved.

\subsection*{Data availability statement}

The data that support the findings of this study are available from the DSS. Restrictions apply to the availability of these data, which were used under license for this study. Data are available at https://dataverse.ada.edu.au/dataset.xhtml?persistentId=doi:10.26193/F2YRL5 with the permission of the DSS. The code used to produce the simulated datasets and carry out the analysis for the simulation study is available on GitHub: https://github.com/rheanna-mainzer/MI-aux-var-selection.

\subsection*{Conflict of interest}

The authors declare no potential conflict of interests.

\bibliography{mybib2}

\end{document}